\documentclass{IEEEtran4PSCC}

\usepackage{scrlfile} 
\PreventPackageFromLoading{fixltx2e}  

\usepackage{cite}

\ifCLASSINFOpdf
   \usepackage[pdftex]{graphicx}
\else
   \usepackage[dvips]{graphicx}
\fi
\usepackage{epstopdf}
\usepackage[dvipsnames]{xcolor}
\usepackage{wrapfig}
\usepackage{pgfplots}
\usetikzlibrary{backgrounds}
\usepgfplotslibrary{colorbrewer}
\pgfplotsset{compat = 1.16, cycle list/Set1-8} 
\usetikzlibrary{pgfplots.statistics} 
\usepackage{pgfplotstable}

\usetikzlibrary{external}
\tikzset{external/force remake=false} 

\pgfplotsset{
    boxplot/draw/average/.code={%
        \color{.!0!black}
        \draw[/pgfplots/boxplot/every average/.try]
            \pgfextra
            \pgftransformshift{%
                \pgfplotsboxplotpointabbox
                    {\pgfplotsboxplotvalue{average}}
                    {0.5}%
            }%
            \pgfuseplotmark{\tikz@plot@mark}%
            \endpgfextra
        ;
    },
}

\definecolor{uugreen}{HTML}{14B03D}
\definecolor{uured}{RGB}{191,45,56}
\definecolor{uublue}{RGB}{0, 106, 178}
\definecolor{backgroundcolor}{rgb}{0.08, 0.38, 0.74}
\definecolor{greenBackgroundcolor}{rgb}{0.4660, 0.6740, 0.1880}
\definecolor{backgroundcolor2}{rgb}{0, 0.4470, 0.7410}

\usepackage{textpos}

\usetikzlibrary{circuits.ee.IEC}
\usetikzlibrary{arrows}
\tikzstyle{roundnode} =[circle, draw=blue!60, fill=blue!5, scale = 0.5]  
\usetikzlibrary{positioning}
\usetikzlibrary{arrows.meta}

\usepackage{environ}
\makeatletter
\newsavebox{\measure@tikzpicture}
\NewEnviron{scaletikzpicturetowidth}[1]{%
  \def\tikz@width{#1}%
  \begin{lrbox}{\measure@tikzpicture}%
  \BODY
  \end{lrbox}%
  \pgfmathparse{#1/\wd\measure@tikzpicture}%
  \BODY
}

\usepackage{dblfloatfix}

\usepackage[cmex10]{amsmath}
\usepackage{amssymb}
\usepackage{amsthm} 
\theoremstyle{definition}

\usepackage{siunitx}
\usepackage{multicol}    
\usepackage{booktabs}  

\makeatletter
\let\old@ps@headings\ps@headings
\let\old@ps@IEEEtitlepagestyle\ps@IEEEtitlepagestyle
\def\psccfooter#1{%
    \def\ps@headings{%
        \old@ps@headings%
        \def\@oddfoot{\strut\hfill#1\hfill\strut}%
        \def\@evenfoot{\strut\hfill#1\hfill\strut}%
    }%
    \def\ps@IEEEtitlepagestyle{%
        \old@ps@IEEEtitlepagestyle%
        \def\@oddfoot{\strut\hfill#1\hfill\strut}%
        \def\@evenfoot{\strut\hfill#1\hfill\strut}%
    }%
    \ps@headings%
}
\makeatother

\usepackage[most]{tcolorbox}

\usepackage{makecell}  
\usepackage{multirow}

\usepackage[T1]{fontenc}  
\usepackage{newtxtext}

\usepackage[caption=false,font=footnotesize]{subfig}

\usepackage[hidelinks]{hyperref}
\usepackage{comment}

\usepackage[bottom]{footmisc}

\usepackage{stackengine}

\input{config/macros}

\begin{document}
%

\title{Ultrafast Grid Impedance Identification in $dq$-Asymmetric Three-Phase Power Systems}


\author{
    Mohamed Abdalmoaty\textsuperscript{*},
    Verena Häberle\textsuperscript{*},
    Xiuqiang He\textsuperscript{$\dagger$},
    and Florian Dörfler\textsuperscript{*}\\
   {\textsuperscript{*}Automatic Control Laboratory, ETH Zurich, 8092 Zurich, Switzerland}\\
   {\textsuperscript{$\dagger$}Department of Automation, Tsinghua University, Beijing 100084, China}
}


\maketitle

\begin{abstract}
We propose a non-parametric frequency-domain method to identify small-signal $dq$-asymmetric grid impedances, over a wide frequency band, using grid-connected converters. Existing identification methods are faced with significant trade-offs: e.g., passive approaches rely on ambient harmonics and rare grid events and thus can only provide estimates at a few frequencies, while many active approaches that intentionally perturb grid operation require long time series measurement and specialized equipment. Although active time-domain methods reduce the measurement time, they either make crude simplifying assumptions or require laborious model order tuning. Our approach effectively addresses these challenges: it does not require specialized excitation signals or hardware and  achieves ultrafast ($< \!1$\! \si{\second}) identification, drastically reducing measurement time. Being non-parametric, our approach  also makes no assumptions on the grid structure. A detailed electromagnetic transient simulation is used to validate the method and demonstrate its clear superiority over existing alternatives.
\end{abstract}

\begin{IEEEkeywords}
Data-driven methods, Equivalent $dq$-impedance, Frequency scan, \!Grid-converter interaction, Small-signal stability \vspace{-0.5cm}
\end{IEEEkeywords}
\vspace{-0.25cm}
\thanksto{\noindent {This work was supported by funds from Elia Group and Energinet via the \textit{2025 Elia Group Research Challenge} on digitalization of system operations.}\vspace{-0.5cm}}

\section{Introduction}\label{sec:into}\vspace{-0.05cm}

With the increasing integration of distributed renewable generation and power-electronics-based technologies, modern power systems are becoming more dynamic. Today's grids exhibit diverse and variable subsystem interactions, making simple analytical models inadequate \cite{Wang2019}. This complexity is compounded by limited data and model sharing among stakeholders: device manufacturers typically withhold proprietary models, grid operators may only have coarse or steady-state system models, and consumers can exhibit significant but largely unknown dynamics (e.g., data centers). In this context, impedance identification offers a promising alternative \cite{DeMeerendre2020}, providing data-driven models of local grid dynamics.

Accurate knowledge of the grid impedance is valuable for a wide range of applications. Its value at the fundamental frequency, $\omega_g$, can be used to estimate voltage stability margins,  maximum power transfer limits,  grid strength metrics,  etc. When characterized over a wide frequency band, it enables harmonic penetration studies and filter design,  characterization of inter-area and subsynchronous oscillations, or model-based control design. It also plays a key role in the optimized operation of grid-connected converters \cite{Harnefors2007a,cespedes2014adaptive,Wang2014,Wang2018}. Interactions between converters and the grid can degrade power quality and trigger instabilities \cite{Mollerstedt2000,Liserre2006,li2017unstable}. Impedance-based analysis has therefore emerged as an effective tool for assessing small-signal stability \cite{belkhayat1997,sun2009small,Rygg2016,Chen2024}, modeling the system’s small-signal dynamics via an equivalent dynamic impedance $Z_g(s)$. In three-phase systems, this impedance forms a multi-input multi-output (MIMO) transfer function relating small-signal terminal voltages and currents at the Point of Common Coupling (PCC), typically represented in a synchronous $dq$-frame. 

During the last two decades, numerous approaches for grid impedance identification have been proposed, primarily in the power electronics literature.  They  can be classified into \emph{passive}, \emph{quasi-passive}, and \emph{active} methods. A detailed description of most methods is given in \cite{Stiegler2015} and \cite{DeMeerendre2020}. 
\emph{Passive} methods rely on the small harmonic distortion that naturally exists at the PCC. Their applicability is limited to impedance estimation at $\omega_g$ and a few harmonic frequencies, \cite{Gu2012,Hoffmann2014}. They may not be suitable for tasks that rely on wideband characterizations, such as stability analysis and advanced control design.  
\emph{Quasi-passive} methods combine a triggering mechanism \cite{Garcia2014,Cobreces2009} with an active method to avoid continuous grid perturbation. \emph{Active} methods deliberately introduce disturbances by repeatedly switching resistive or capacitive loads \cite{Girgis1989,Jordan2018} or by injecting small current or voltage signals through dedicated hardware \cite{francis2011algorithm,huang2009small,Rygg2016}. Examples include frequency-sweep techniques \cite{huang2009small}, which apply perturbations sequentially at discrete frequencies, and wideband excitation methods \cite{Rygg2016}, which excite multiple frequencies simultaneously. Both use idealized steady-state Fourier analysis of the measured voltages and currents to obtain a non-parametric frequency-domain model. However, they require specialized hardware and long measurement times, which limits their practicality.

A more practical approach is to use existing grid-connected converters to excite the grid when needed by superimposing an excitation signal on the reference of an inner control loop. Impulse excitations have been proposed \cite{cespedes2012online,liu2020analysis}, offering very short perturbation times but at the cost of large disturbances that may jeopardize power quality, excite nonlinear behavior, or trigger protection relays. Alternatively, smaller-amplitude signals such as Maximum-Length Binary Sequences (MLBS) have been employed \cite{martin2013wide,riccobono2017noninvasive,luhtala2018implementation,roinila2017mimo}; however, these methods either assume $dq$-symmetry and neglect cross-coupling effects  (i.e., a diagonal $Z_g(s)$), or require sequential, linearly independent perturbations, resulting in longer measurement times.

Despite considerable progress, accurate grid impedance identification remains challenging due to inherent trade-offs. A key difficulty is minimizing the perturbation time and amplitude while preserving the accuracy of the identification. Recent efforts have sought to address this: for instance, \cite{Berg2022} proposed a non-parametric frequency-domain approach using orthogonal binary signals as an alternative to sequential perturbation. Although potentially more robust, its measurement time is the same as MLBS-based methods \cite{martin2013wide,riccobono2017noninvasive,luhtala2018implementation,roinila2017mimo} and still requires periodic steady-state measurements to prevent spectral leakage in discrete Fourier transform (DFT) analysis. Alternatively, \cite{Haberle2023} demonstrated the use of parametric time-domain techniques, particularly discrete-time Auto-Regressive Exogenous (ARX) models \cite{ljung1998system}, which eliminate the need for specialized excitation signals and can handle non-periodic or transient measurements. However, their performance is rather sensitive to the chosen parameterization and model order.

To address these challenges, we propose an active \emph{non-parametric} frequency-domain identification method for $dq$-\textit{asymmetric} grids that does not require sequential perturbations or steady-state measurements. This eliminates the measurement-time limitations of existing methods, leaving only constraints imposed by the required frequency resolution and signal-to-noise ratio (SNR). We leverage complex  TFs \cite{Harnefors2007} to parametrize the grid equivalent impedance using single-input single-output (SISO) complex TFs and show how to reconstruct the full real grid impedance TF matrix once an estimate is obtained. This approach simplifies MIMO impedance parameterization, clearly distinguishes between symmetric and asymmetric cases, and provides an algebraically efficient representation. Furthermore, the SISO complex TF non-parametric estimates can be directly used for stability assessment or control design \cite{Zhang2018,Harnefors2020}. They may also be converted to parametric models using vector fitting methods \cite{Ozdemir2017}.

The proposed approach relies on frequency-domain local parametric approximations that are well studied in the system identification literature \cite{Pintelon2010,McKelvey2012,Pintelon2021}. We assume that the frequency response of the complex TFs can be accurately approximated over short frequency intervals by low-order continuous-time ARX models. This is a reasonable assumption, in particular, when the number of samples $N$ ensures that the 3dB-bandwidth of any resonance spans several spectral lines. No finite global order is assumed, making the estimated model truly non-parametric. Local models are fitted by solving $N$ small and independent linear least-squares problems. As shown in simulations, the identification accuracy is insensitive to the local model order, unlike parametric methods.

In summary, our approach strikes a balance between fully non-parametric methods that require sequential perturbations, and fully parametric methods that attempt capturing the dynamics with a single high-order parametric model, often with insufficient accuracy. 
It offers ultrafast ($<\! 1$\! \si{\second}) identification with a favorable  accuracy and data efficiency trade-off.
\section{Problem formulation}\label{sec:formulation}

\subsection{Small-signal model}
The objective is to identify the dynamic small-signal Th\'{e}venin equivalent impedance of an AC three-wire, three-phase grid. This is achieved using time-domain samples of terminal voltages $v_{abc}$ and currents $i_{abc}$ at the PCC of interest; see Figure \ref{figure:fig1}. No assumptions are made about the topology or strength of the grid, which can include generators, loads, and actively controlled power-electronics systems. Under balanced operation, Park’s transformation at the steady-state frequency of the grid $\omega_g$ maps the three-phase voltages and currents to constant quantities in synchronous $dq$-coordinates, providing a steady-state operating point for small-signal linearization.

The small-signal grid impedance model is  given by four SISO real transfer operators that relate the $dq$ small-signal currents and voltages,\vspace{-0.4cm}
\[
\begin{bmatrix}
    \Delta v_d(t)\\
    \Delta  v_q(t)
\end{bmatrix} =
\overbrace{\begin{bmatrix}
    Z_{dd}(\p) & Z_{dq}(\p)\\
    Z_{qd}(\p) & Z_{qq}(\p)\\
\end{bmatrix}}^{=:Z_g(\p)}
\begin{bmatrix}
    \Delta  i_d(t)\\
    \Delta  i_q(t)
\end{bmatrix},
\]
with $Z_g(\p)$ being real 2-by-2 transfer operators,  $\p\!=\!\frac{\diff}{\diff t}$ is the differential operator, and $\Delta$ denotes deviations from steady-state values.\footnote{In practice we get these by removing the mean of the time series \vspace{-0.55cm}} An alternative equivalent representation can be obtained using complex variables. Define
\[
 \bv(t) := \Delta v_d(t) +j \Delta v_q(t),  \qquad \bi(t) := \Delta i_d(t) + j \Delta i_q(t).
\]
Straightforward algebraic manipulations \cite{Martin2004}  then show that  
\begin{equation}\label{eq:complex_model}
\bv(t) = \bG_+(\p) \bi(t) + \bG_-(\p) \bi^\ast(t),
\end{equation}
where $\bi^\ast$ is the complex conjugate of $\bi$, and
\begin{equation}\label{eq:complex_tfs}
\begin{aligned}
   \bG_+(\p) &=  0.5 \big({Z_{dd}(\p) + Z_{qq}(\p)} + j ({Z_{qd}(\p)  - Z_{dq}(\p)}) \big)   \\
   \bG_-(\p) &= 0.5 \big({Z_{dd}(\p) - Z_{qq}(\p)} + j({Z_{dq}(\p)  + Z_{qd}(\p)})\big)
\end{aligned}
\end{equation}
are two SISO complex transfer operators. TFs are obtained by applying the Laplace transform to the time-domain transfer operators, under which  $\p$ becomes the Laplace variable $s$. The frequency response at $\omega$ is obtained by setting $s = j\omega$.

If $Z_g(\p)$ is symmetric, i.e. $Z_{dd}(\p) = Z_{qq}(\p) = G_d(\p)$ and $Z_{qd}(\p) = -Z_{dq}(\p) = G_q(\p)$, it holds that $\bG_-(\p) = 0$ and the model reduces to $\bv(t) = \bG(\p) \bi(t)$,
with
\begin{equation}
\label{eq:single_complex_tf}
\bG(\p) = G_d(\p) + j G_q(\p) = \bG_+(\p).
\end{equation}
In this special case, $Z_g(\p)$ is represented by one SISO complex TF that can be estimated non-parametrically via one set of periodic measurements. However, in the asymmetric case, $\bi^\ast$ is always required, resulting in a double frequency model that cannot be estimated using one set of measurements unless further assumptions are imposed on the frequency response.

\subsection{Grid-connected converter}
\begin{figure}
    \centering
\resizebox {0.9\linewidth} {!} {
    \includegraphics{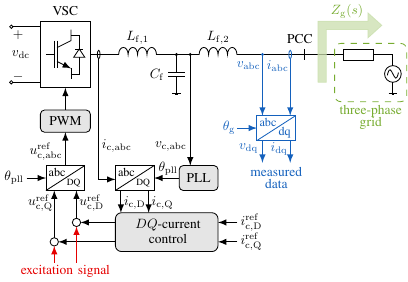}}
        \vspace{-.3cm}
    \caption{Grid-connected converter system with excitation in the control loop.} \vspace{-0.3cm}
    \label{figure:fig1}
\end{figure}

To demonstrate the approach, we consider a grid-connected voltage-sourced converter with an $LCL$ filter, as shown in Figure \ref{figure:fig1}. An ideal DC link is assumed. The current control loop is implemented in the $dq$-frame, and grid synchronization is performed using a PLL that tracks the voltage on the $LCL$ filter capacitor. The current controller output serves as the reference for the converter voltage, realized via PWM. Outer power control loops are omitted for the sake of clarity.

The grid is perturbed by adding a wideband excitation signal to the converter voltage references $u_\mathrm{c,D}^\mathrm{ref}$ and $u_\mathrm{c,Q}^\mathrm{ref}$. This provides a wider excitation bandwidth than adding it to the current reference. We use a zero-mean random binary signal (RBS) \cite{ljung1998system}, which, unlike MLBS, is non-periodic and can have arbitrary length. Binary signals are popular due to their ideal crest factor, but they are inflexible for spectrum shaping. Excitation spectrum design is beyond the scope of this work.

\subsection{Measurement setup}
\begin{figure}
    \centering
\resizebox {0.88\linewidth} {!} {
	 \includegraphics{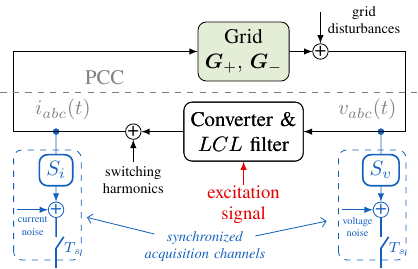}
	}
	\vspace{-0.1cm}
    \caption{Measurement setup. Voltage and current noise represent errors due to inaccuracies of the measurement devices $S_i$ and $S_v$. Grid disturbances represent possible ambient harmonics and/or transient events in the grid. The natural converter's switching harmonics act as an additional excitation signal.} \vspace{-0.2cm}
    \label{figure:measurment_setup}
\end{figure}

The measurement setup is shown in Figure \ref{figure:measurment_setup}. Samples of voltage $v_{abc}(t)$ and current $i_{abc}(t)$ are recorded, starting at the application instant of the excitation signal, and continue for its entire duration.   For accurate identification, $v_{abc}(t)$ and $i_{abc}(t)$ are filtered with an anti-alias filter, such as a Chebyshev analog filter, before sampling. In addition, acquisition channels must be synchronized and relatively calibrated to eliminate transducer dynamics ($S_i$ and $S_v$). Sampling is uniform in time (periodic) and should be fast enough for the bandwidth of interest; a possibility is to use the converter's sampling time $T_s$, which typically operates at the switching frequency or twice that (tens of \si{\kilo\hertz}). Lastly, Park’s transformation converts the sampled phase measurements to a synchronous $dq$-frame.
\section{Grid Impedance Identification}\label{sec:analysis}

Given a data set \mbox{$D_N \!:= \!\{ (\bv(t_n), \bi(t_n)),  n\! \in\! \{0, \dots, N\!-\!1\} \}$}, the  goal is to construct a non-parametric estimator of $\bG_+(j\omega), \bG_-(j\omega)$ on a uniform grid of frequencies $\omega_k\! =\! \frac{2\pi k}{NT_s}$, viz. $D_N \, \mapsto \, \{ (\, \widehat{\bG_+(j\omega_k)},\; \widehat{\bG_-(j\omega_k)}\,), \;\, k \!\in\!\{ 0, \dots, N-1\} \}$.
To this end, let $ \bV\!_k $ and $ \bI_k $ denote the $N$-point DFT of $ \{ \bv(t_n) \} $ and $ \{ \bi(t_n) \} $, respectively, defined as \vspace{-0.05cm}
\begin{equation}\label{eq:dft_spectrum}
\bV\!_k = \frac{1}{\sqrt{N}} \sum_{n=0}^{N-1} \bv(t_n) \eu^{-j\omega_k n T_s}, \quad k \in \{0, \dots, N-1\}, \vspace{-0.05cm}
\end{equation}
 and a similar expression for $ \bI_k $. Recall that the $N$-point DFT can be computed very efficiently using fast Fourier transform algorithms, and that the spectra are periodic with period $N$, namely $\bI_{k+N} = \bI_{k}$. Because we are dealing with complex time-domain signals, the Hermitian symmetry property of the DFT does not hold, but it still satisfies conjugacy and reversal properties, that is, the DFT of $\{\bi^\ast(t_k)\}$ is given by   $\{\bI^\ast_{(N-k)_N}  \}$ where $(N-k)_N$ stands for ``$N-k$ modulo $N$".

\subsection{Relation between voltage and current DFT spectra}

Applying the finite-time (truncated) Fourier transform to \eqref{eq:complex_model} over $[0,NT_s]$, gives the model
\begin{equation}\label{eq:dft_relation}
\begin{aligned}
\bV\!_k &= \bG_+(j\omega_k) \bI_k + \bG_-(j\omega_k) \bI_{(N-k)_N}^\ast + \bT(j\omega_k),\\
\end{aligned}
\end{equation}
where $ \bT(j\omega_k) $ is a transient term decaying at a rate $ \mathcal{O}({N^{\frac{1}{2}}})$. It accounts for sampling, spectral leakage which arises from the mismatch in initial and final conditions, and aliasing effects due to the truncations to the finite time interval $[0, NT_s]$. We emphasize  that \eqref{eq:dft_relation} is an exact relation between the DFT spectra even for arbitrary non-periodic measurements. For more details, the interested reader is referred to \mbox{\cite[Thm. 2.1]{ljung1998system},} \cite{Pintelon1997} and the Appendix. 

This shows that identifying the grid impedance from a single \textit{short} measurement cycle using DFT spectra presents two main challenges. The first stems from the error introduced by the transient term $\bT$. The second is that there are $2N$ complex unknowns, but only $ N $ complex data equations are available. The first challenge is solved by estimating $\bT$ together with the complex TFs. The second challenge is addressed by using local parametric modeling, which provides additional data equations by using neighboring spectral lines, as outlined below.

\subsection{Local parametric modeling}
The idea of local parametric modeling (see \cite{Pintelon2010, McKelvey2012, Pintelon2021}) is as follows. To obtain a non-parametric estimate at a frequency $\omega_k$, we approximate the complex TFs $\bG_+(j\omega_k), \bG_-(j\omega_k)$, and $\bT(j\omega_k)$ over a short frequency range around $\omega_k$ using a parametric model of low order. Estimates at different frequencies are correlated only via raw data, and therefore the method remains truly non-parametric in nature.

By their definition in \eqref{eq:complex_tfs}, $\bG_+$ and $\bG_-$ have the same poles, which are also poles of $\bT$. Therefore, for a local frequency interval $[\omega_{k-\ell}, \omega_{k+\ell}] $ centered on $\omega_k$, we approximate
\begingroup\makeatletter\def\f@size{9.5}\check@mathfonts
\begin{equation}\label{eq:LPM}
\begin{aligned}
        \bG_+(j\omega_{k+r}) &\approx \frac{\bB_k^+(j\omega_{r})}{\bA_k(j\omega_{r})}, \quad \bG_-(j\omega_{k+r}) \approx \frac{\bB_k^-(j\omega_{r})}{\bA_k(j\omega_{r})}\\
          \bT(j\omega_{k+r}) &\approx \frac{\bC_k(j\omega_{r})}{\bA_k(j\omega_{r})},
\end{aligned}
\end{equation}
\endgroup
where $r \in \{-\ell, \dots, \ell\}$, $\bA_k, \bB^+_k, \bB^-_k$, and  $\bC_k$ are defined as complex polynomials in $j\omega_r$ of degree $R$, parameterized as $\bA_k(j\omega_r) :=  \sum_{m = 0}^R a_m(k) r^m$, and similarly for the other polynomials.  Although these polynomials can have different degrees, using the same fixed degree for all and keeping it constant over $k$ provides a simple and effective choice. We normalize $\bA_k$ so that $a_0(k) = 1$ for every $k$. With this parameterization, the vector of parameters to be estimated is  \vspace{-0.1cm}
\begin{align*}
&\theta_k :=\left[\begin{matrix}
    a_1(k) \cdots a_R(k) & b_0^+(k)    \cdots\end{matrix}\right. \\
& \hspace{2cm}  \left.\begin{matrix} \cdots b_R^+(k) & b_0^-(k)\cdots  b_R^-(k) & c_0(k) \cdots c_R(k) \end{matrix}\right]. 
\end{align*}

\vspace{-0.05cm}
\noindent It contains $4R+3$ unknown complex parameters. From \eqref{eq:dft_relation} and \eqref{eq:LPM}, the local model linking the DFT spectra becomes\vspace{-0.05cm}
\[
\begin{aligned}
    \bA_k(j\omega_r) \bV\!_{k+r} = \,& 
    \bB_k^+(j\omega_r) \bI_{k+r} 
    + \bB_k^-(j\omega_r) \bI^\ast_{(N-k-r)_N}\\& + \bC_k(j\omega_r) + \bE_{k+r}
\end{aligned}\vspace{-0.05cm}
\]
where $\bE_{k+r}$ accounts for the local parametric interpolation error. The parameters for each $k$ can then be estimated by minimizing the errors in the least squares sense.\footnote{This corresponds to fitting a \textit{local continuous-time} ARX model to the spectra only over $[\omega_{k-\ell}, \omega_{k+\ell}]$.\vspace{-.5cm}} 

Define the complex column vectors and the data matrix\vspace{-0.05cm}
\[
\begin{aligned}
\bY\!_k &:= \begin{bmatrix}
\bV\!_{k-\ell}\!\! &\dots &\!\!\bV\!_{k+\ell}
\end{bmatrix}^\top\!\!, \quad 
\bU^+_k := \begin{bmatrix}
\bI_{k-\ell} \!\!&\dots\!\! &\bI_{k+\ell}
\end{bmatrix}^\top\!\!,\\
\bU^-_k &:= \begin{bmatrix}
\bI^\ast_{(N-k-\ell)_N} &\dots &\bI^\ast_{(N-k+\ell))_N}
\end{bmatrix}^\top,\\
\Phi_k &:= \begin{bmatrix}
    \bY\!_k  \otimes \tilde{\phi}(r) & 
    \bU^+_k \otimes \phi(r) & 
    \bU^-_k \otimes \phi(r) & 
   \1 \otimes  \phi(r) 
\end{bmatrix},
\end{aligned}\vspace{-0.05cm}
\]
with $\tilde{\phi}(r) = \begin{bmatrix} r & r^2&  \dots& r^R\end{bmatrix}^\top$ and $\phi(r) := \begin{bmatrix}1 &\tilde{\phi}(r)^\top\end{bmatrix}^\top$, where $\otimes$ denotes the Kronecker product. The parameter estimate is then obtained by solving the linear least-squares problem $\min_\theta\| \bY\!_k - \Phi_k \theta  \|$, with the closed-form solution
\begin{equation}\label{eq:LS_sol}
\hat{\theta}_k = \Phi_k^\dag \bY\!_k.
\end{equation}
The matrix $\Phi_k^\dag$ denotes the pseudo-inverse of $\Phi_k$, which may be computed using a singular value decomposition with an appropriate scaling to improve numerical conditioning. The frequency response estimates at $\omega_k$ are   computed by setting $r=0$ in \eqref{eq:LPM} and recalling that $a_0(k) = 1$. This gives $\widehat{\bG_+(j\omega_{k})} = \widehat{b_0^+(k)}, \quad \widehat{\bG_-(j\omega_{k})} = \widehat{b_0^-(k)}$. 
Notice that the least-squares problems are independent over $k$, so the computations may be optimized by parallelization.

To guarantee the uniqueness of \eqref{eq:LS_sol},   the local frequency interval should be wide enough, and the measurement should be sufficiently exciting. A necessary condition is $2\ell\!  +\!1 \geq 4R\! +\! 3$ where $\ell$ is the radius of the local frequency interval.  The measurement is sufficiently exciting at  $\omega_k$ if $\Phi_k$ is full rank. This imposes a condition on the local spectra $\bY_k$ and $\bU_k$, and is satisfied for an RBS excitation signal. Notably, the proposed approach can estimate the equivalent impedance at $\omega_g$ because it employs local models. This is despite the fact that the RBS does not excite the grid at $\omega_g$.

\subsection{Special cases}\vspace{-0.05cm}

\subsubsection{The periodic excitation case}
If the excitation signal is repeated periodically and $D_N$ contains an integer number of steady-state periods, then $\bT(j\omega) = 0$. In this case, it is not necessary to estimate the polynomials $\bC_k$, and the corresponding columns in $\Phi_k$ can be safely removed. However, local parametric modeling is still needed due to the lack of a second measurement.

\subsubsection{The \texorpdfstring{$dq$}{dq}-symmetric grid case}
If it is known a priori that the equivalent impedance is $dq$-symmetric, the model reduces to
$\bV\!_k = \bG(j\omega_k) \bI_k + \bT(j\omega_k)$.
In this case, it is not necessary to estimate the polynomials $\bB_k^-(j\omega)$, and the corresponding columns in $\Phi_k$ can be safely removed.
Although here we only have one SISO complex TF to identify, the effect of leakage errors remains. Therefore, local parametric modeling is still needed to eliminate the spectral leakage.

\subsubsection{The periodic excitation and \texorpdfstring{$dq$}{dq}-symmetric grid case}
Without spectral leakage or $dq$-asymmetry, an estimate of $\bG$  at frequency $\omega_k$ can be obtained by a single division $\widehat{\bG(j\omega_k)} = {\bV\!_k}/{\bI_k}$. However, the use of this simple estimator (a.k.a empirical transfer function estimate, ETFE \cite[(6.24)]{ljung1998system}) is justified only if steady-state measurements are possible; otherwise, significant errors would occur \cite[Lem. 6.1]{ljung1998system}.

\subsection{Extracting the four real TFs}\vspace{-0.05cm}
As pointed out earlier, non-parametric estimates of complex TFs can be used directly for control and stability analysis. However, if needed, they can be mapped numerically to non-parametric estimates of $Z_g$. For $\bG(s)$ in \eqref{eq:single_complex_tf}, by
noticing that  $\bG^\ast(s)  =  [\bG(s^\ast)]^\ast$,
we find that  $G_d(s) = 0.5({\bG(s) + \bG^\ast(s)})$,  $G_q(s)\! =\! -0.5j({\bG(s) \!-\! \bG^\ast(s)})$. Hence,  in the symmetric case\vspace{-0.05cm}
\begingroup\makeatletter\def\f@size{9}\check@mathfonts
\[
\begin{aligned}
 \widehat{Z_{dd}(j\omega_k)}  &= \frac{1}{2} (\widehat{\bG(j\omega_k)} + [\widehat{\bG(j\bar{\omega}_k)}]^\ast),\\  
 \widehat{Z_{qd}(j\omega_k)} &= \frac{1}{2j} (\widehat{\bG(j\omega_k)} - [\widehat{\bG(j\bar{\omega}_k)}]^\ast), \vspace{-0.1cm}
\end{aligned}
\]
\endgroup
where $\widehat{\bG(j\bar{\omega}_k)}$ is the estimate of $\bG(-j\omega_{k})$,  $\bar{\omega}_k = \omega_{(N-k)_N}$, and  $k \in\{0,\dots, \frac{N}{2}\! -\!1\}$ assuming an even $N$. Using a similar reasoning for the asymmetric case, \eqref{eq:complex_tfs} gives \vspace{-0.05cm}
\begingroup\makeatletter\def\f@size{8}\check@mathfonts
\[
\begin{aligned}
\widehat{Z_{dd}(j\omega_k)} &= \frac{1}{2}   \Big( \widehat{\bG_+(j\omega_k)} + [\widehat{\bG_+(j\bar{\omega}_k)}]^\ast + \widehat{\bG_-(j\omega_k)} + [\widehat{\bG_-(j\bar{\omega}_k)}]^\ast \Big),\\
\widehat{Z_{qq}(j\omega_k)} &=  \frac{1}{2}  \Big( \widehat{\bG_+(j\omega_k)} + [\widehat{\bG_+(j\bar{\omega}_k)}]^\ast - \widehat{\bG_-(j\omega_k)} - [\widehat{\bG_-(j\bar{\omega}_k)}]^\ast \Big),       \\
\widehat{Z_{dq}(j\omega_k)} &= \!\frac{-1}{2j} \Big( \widehat{\bG_+(j\omega_k)}\! - [\widehat{\bG_+(j\bar{\omega}_k)}]^\ast\! - \!\widehat{\bG_-(j\omega_k)} + [\widehat{\bG_-(j\bar{\omega}_k)}]^\ast \Big),\\
\widehat{Z_{qd}(j\omega_k)} &= \frac{1}{2j}  \Big( \widehat{\bG_+(j\omega_k)}\! - [\widehat{\bG_+(j\bar{\omega}_k)}]^\ast \!+ \!\widehat{\bG_-(j\omega_k)} - [\widehat{\bG_-(j\bar{\omega}_k)}]^\ast \Big).
\end{aligned}
\]
\endgroup

\section{Numerical simulation study}\label{sec:simulations}
\subsection{Experiment Setup}
We demonstrate the performance of the proposed approach using detailed electromagnetic transient simulations in Matlab/Simulink using the Simscape Electrical toolbox.\footnote{Code is available at \url{https://doi.org/10.3929/ethz-c-000784827}.\vspace{-0.5cm}} The converter is modeled via an IGBT bridge and a discrete space-vector PWM.  The \texttt{ode5} solver is used with a  step size $10^{-7} \si{\second}$.
  
\begin{figure}[t]
    \centering
\resizebox {0.49\textwidth} {!} {
        \includegraphics{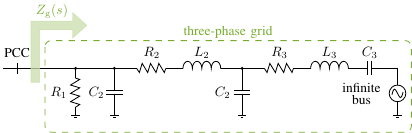}
        }
    \vspace{-0.6cm}
    \caption{One-line diagram of the three-phase grid used in the simulation.}
    \label{fig:simulation_grid}
    \vspace{-0.15cm}
\end{figure}

\begin{figure}[ht]
    \centering
    \includegraphics{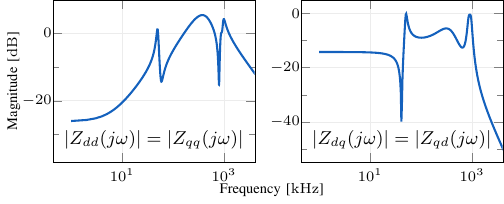}
     \vspace{-0.3cm}
     \caption{Magnitude frequency response of the true equivalent impedance $Z_g(s)$} \vspace{-0.35cm}
     \label{fig:true_model}
\end{figure}

Figure \ref{fig:simulation_grid} shows the one-line diagram of the grid. Its analytically computed equivalent impedance exhibits rich dynamics with a few resonances, as shown in Figure \ref{fig:true_model}. The converter, interfaced with the grid using an $LCL$ filter,  is controlled as shown in Figure \ref{figure:fig1}, in addition to an outer PQ loop. The controllers are discretized parallel-form proportional–integral (PI) regulators. All relevant parameters are listed in Table \ref{tab:experiment_parameters}.

\renewcommand{\arraystretch}{1.1}
\begin{table}[b!]\scriptsize
    \centering
    \vspace*{-0.5cm}
           \caption{Parameters of the Simulated Grid/Converter}
           \vspace{-0.2cm}
           \begin{tabular}{c||c|c}
        \toprule
         Parameter & Symbol & Value  \\ \hline
         $\hspace{-1mm}$Voltage, power \& freq. base$\hspace{-1mm}$& $V_\mathrm{b},\,S_\mathrm{b},\,f_\mathrm{b}$ & 380 \si{\volt}, 1.5 kVA, 50 Hz\\
         LCL filter components& $\hspace{-1mm}$$L_\mathrm{f,1},\,L_\mathrm{f,2},\,C_\mathrm{f}$$\hspace{-1mm}$& 0.08 p.u.\, 0.05 p.u., 0.08 p.u.\\
         Load component& $R_1$ & 2 p.u.\\
         Line 1 components& $R_2,\,L_2,\,C_2$ & 0.015 p.u., 0.15 p.u., 0.05 p.u.\\
         Line 2 components& $R_3,\,L_3,\,C_3$ & 0.015 p.u., 0.15 p.u., 10 p.u.\\
         DC link voltage & $v_dc$ & 1150 \si{\volt}\\\hline
         Current PI controller gains &  \multicolumn{2}{l}{$k_i = 10 \text{ p.u.}, \hspace{1.5cm}  k_p = 0.3 \text{ p.u.}$ }\\
         PQ PI controller gains & \multicolumn{2}{l}{$k_i = 40 \text{ p.u.},\hspace{1.5cm}  k_p = 0.5 \text{ p.u.}$} \\
         PLL PI controller gains &   \multicolumn{2}{l}{$k_i = 40^2/(2+\sqrt{5}), \;\;\; k_p = \sqrt{2k_i}$ \;\si{\radian\per\second} } \\
         Converter's set point & \multicolumn{2}{l}{$P = 0.8,\;  Q = 0$ p.u.} \\
         Switching/control frequency &  \multicolumn{2}{l}{ 10 \si{\kHz}}\\\hline
	Equivalent impedance $Z_g(s)$  &  \multicolumn{2}{l}{ $dq$-symm. rational $2 \!\!\times\!\! 2$  TF matrix of (order = 10)}	\\
         \bottomrule
    \end{tabular}\vspace{-0.25cm}
     \label{tab:experiment_parameters}
\end{table}
\renewcommand{\arraystretch}{1}\normalsize

For clarity, we assume stationary grid operating conditions and do not consider any grid ambient harmonics (no grid disturbances). However, we note that the approach can deal with ambient harmonics efficiently by removing the corresponding spectral lines from the identification process.  Also note that although the grid  $dq$-symmetric, the methods do not assume this symmetry. Instead, they identify a 2-by-2 MIMO model.

To ensure that the validation scenario remains close to a realistic one, the experiment parameters are first fixed to reflect practical limitations. In particular, the total excitation time is fixed at 1 \si{\second} (and only one measurement cycle), and the sampling time  $T_s =10\,\si{\milli\second}$ is equal to the converter's switching/control period. The amplitude of the excitation signal is $\leq  0.05$ p.u. resulting in an acceptable $\text{THD}_v$ ($\approx 2.15\%$)\footnote{Based on 10 cycles,  all intraharmonics  \& up to the 50\textsuperscript{th} harmonic.\vspace{-0.5cm}} that does not severely compromise power quality. We consider measurement devices of accuracy class 0.5\% and model current and voltage noise as discrete-time Gaussian random variables. Note that in this scenario, $N = 10^4$ and the DFT uniform frequency grid has a resolution of 1 \si{Hz}, which allows resonance peaks with a 3\si{\dB}-bandwidth of 5 \si{\Hz} to be captured by 5 spectral lines.

\subsection{Comparison methods}
Most existing grid impedance identification methods would struggle to provide an accurate estimate in this scenario, mainly due to the short measurement time. Approaches such as frequency sweep and impulse injection methods are not even applicable, as they require multiple measurement cycles and large excitation amplitudes, respectively. While other wideband DFT-based methods (e.g.\cite{martin2013wide,riccobono2017noninvasive,luhtala2018implementation,roinila2017mimo,Berg2022}) can be applied, their accuracy is severely compromised by the short measurement time. Discrete-time parametric methods (e.g., \cite{Haberle2023}), in contrast, are capable of handling short non-periodic measurements, but their accuracy is highly sensitive to the model order. 

The proposed approach provides a robust solution to these inherent challenges. To demonstrate this, we compared the proposed approach with a parametric time-domain method using ARX models \cite{Haberle2023} and a non-parametric sequential perturbation method similar to \cite{martin2013wide,riccobono2017noninvasive,luhtala2018implementation,roinila2017mimo,Berg2022}. For the latter, i) the data set $D_N$ is divided into two equal parts and treated as two different measurements, and ii) the standard Hamming window \cite{ljung1998system} is applied before computing the  DFT spectra to reduce spectral leakage. The proposed approach is tested using local model orders $2, 4, \dots, 10$, and the radius of the local frequency interval $\ell = 4R+2$. The parametric discrete-time ARX method is tested using model orders (output lags) $2, 4, \dots, 10$ and $20$.  

\subsection{Evaluation}\vspace{-0.05cm}
To isolate errors caused by measurement noise from those coming from the methods themselves, we simulated two cases: with and without measurement noise. The identification accuracy is reported using the following metrics.
For any of the real TFs, let  $Z_{0:k} := \begin{bmatrix} Z(j\omega_0) & Z(j\omega_1) & \dots & Z(j\omega_k)  \end{bmatrix}^\top$ be a vector of the true TF evaluated at frequencies $\omega_0$ to $\omega_k$, and denote its estimate by $\widehat{Z}_{0:k}$; the fit metric is then defined as
\begingroup\makeatletter\def\f@size{9.5}\check@mathfonts
\begin{equation}\label{eq:fit}
\text{Fit\%} := 1 - \frac{\|\widehat{Z}_{0:k} -  Z_{0:k} \|_2^2}{\| Z_{0:k} - \text{mean}(Z_{0:k})   \|_2^2} \times 100,
\end{equation}
\endgroup
with $\text{mean}(Z_{0:k}) = \frac{1}{k+1} \sum_{n = 0}^k Z(j\omega_n)$. 
Notice that \text{Fit\%} can assume negative values; larger values indicate better estimates.  A perfect estimate has a fit of 100\%. We also consider the following metric defined for the estimates $\{\!\widehat{Z_g(\omega_\ell)}\!\}$:\vspace{-0.1cm}
\begingroup\makeatletter\def\f@size{9.5}\check@mathfonts
\begin{equation}\label{eq:hinferror}
\text{relative } H_\infty  \text{ error }  := \frac{\max\limits_{\ell\in \{0, \dots, k\}} \bar{\sigma}\left(\widehat{Z_g(j\omega_\ell)} - Z_g(j\omega_\ell) \right)}{ \max\limits_{\ell\in\{0, \dots, k\}}\bar{\sigma}\left(Z_g(j\omega_\ell)\right)},\vspace{-0.05cm}
\end{equation}
\endgroup
$\{\!Z_g(j\omega_\ell)\!\}$ are the true values,  $\bar{\sigma}$ is the largest singular value.

\subsection{Comparison results without measurement noise}\vspace{-0.05cm}

The results for the case without measurement noise are summarized in Table \ref{tab:comparison_without_noise}, where the two accuracy metrics \eqref{eq:fit} and \eqref{eq:hinferror} are evaluated in the frequency band $[0,4]$ kHz (i.e. up to 80\% of the Nyquist frequency). In this case, the error originates from the method itself; thus, a reliable identification method should exhibit high accuracy.

As the results show, the proposed approach gives an almost perfect estimate for all local orders $R = 2, 4, \dots, 10$. In contrast, the discrete-time ARX method has large errors for small orders; these are model misspecification errors. In this example, they become negligible if high model orders are used. Recall that here the order of  $Z_g(s)$ is 10; however, in practice  a true finite order may not exist. We also remark that the models obtained by the discrete-time parametric ARX method are not always stable. Lastly, the sequential perturbation method incurs significant errors, as expected,  due to the spectral leakage despite the use of a Hamming window.

\setlength{\tabcolsep}{3pt}
\renewcommand{\arraystretch}{1.1}
\begin{table}[b!]\scriptsize
    \centering
    \vspace{-3mm}
           \caption{Accuracy comparison {\textit{\underline{without}}} measurement noise\\over frequency band $[0,4]$ \si{\kHz}}
           \vspace{-0.2cm}
               \begin{tabular}{c c || cccc||c}
     \toprule
      \multicolumn{2}{c||}{\makecell{Method\\ (excitation time = 1 \si{\sec})}} & \multicolumn{4}{c||}{\makecell{Fit\%\\(rounded to a single digit)} } &   \multirow{2}{*}{\makecell{Relative\\ $H_\infty$ error}}\\
      
          \multicolumn{2}{c||}{RBS amplitude $\pm$ 0.05 p.u.} & $Z_{dd}$ &  $Z_{dq}$ & $Z_{qd}$ & $Z_{qq}$  &    \\ \hline
                 \multicolumn{1}{c}{~}      &  \makecell{\rule{0pt}{2.5ex}order}  & & & &  \\[0.5em]
         \multicolumn{1}{c|}{\makecell{\color{backgroundcolor} \textbf{Our approach} \\  \color{backgroundcolor} (non-parametric) }}    &\color{backgroundcolor}\bf  2,4,...,10 & \multicolumn{4}{c||}{\makecell{\color{backgroundcolor}\textbf{rounded Fit\% = 100 for all}}} & \color{backgroundcolor} $\bf{< \!3\!\!\times\!\! 10^{-3}}$ \\[-0.3em]
                &&&&&\\
                \multicolumn{1}{c|}{\multirow{3}{*}{\makecell{ discrete-time ARX\\ (parametric) }}}      &  2  & 58.3  & 17.3& 20.2 & 62.2  & 0.9144  \\
                \multicolumn{1}{c|}{~}                        &  4  & 35.7 & -11.6 & -15.7  & 40.6 & 0.9781\\
                \multicolumn{1}{c|}{~}                        &  6  & 98.8 & 96 & 95.9 & 98.9  & 0.9790\\
                \multicolumn{1}{c|}{~}                        &  8  & 99.4 & 98.2 & 98.2  & 99.4 & 0.6616\\
                \multicolumn{1}{c|}{~}                        & 10 & 100 & 100 & 100 & 100      & 0.0619 \\
                \multicolumn{1}{c|}{~}                        & 20 & 100 & 100 & 99.8 & 100     & 0.3809\\   
                &&&&&\\
                \hline
         \multicolumn{2}{c||}{\makecell{\rule{0pt}{3ex} Sequential \\ perturbation } }     & -0.5490  & -2.9673  & -2.7348 &  -0.5504 &  20.3387\\[-1em]
          \multicolumn{2}{c||}{~ }    &    \multicolumn{4}{c||}{\makecell{all values $\times 10^{6}$\\(significant leakage errors)}} &\\
         \bottomrule
    \end{tabular}
     \label{tab:comparison_without_noise}
\end{table}
\renewcommand{\arraystretch}{1} \normalsize
\setlength{\tabcolsep}{6pt}

\subsection{Comparison results with measurement noise}

The results for the case with measurement noise are summarized in Table \ref{tab:comparison_with_noise}, where the two accuracy metrics \eqref{eq:fit} and \eqref{eq:hinferror} are evaluated in the frequency band $[0,2]\, \si{\kHz}$. Because the sequential perturbation method proved inadequate, we  only compare the other two methods. The errors in this case originate from two sources: measurement noise and systematic errors of the methods.

The errors of the discrete-time parametric ARX method are relatively large for low orders. In fact, they are comparable to the errors obtained with noise-free data. This indicates that for low orders, the misspecification errors dominate the noise errors. Only with high order models the fit becomes acceptable. Yet, the relative $H_\infty$ error of the model with the best Fit\% (order 20) remains large (about 91\%). This is due to the large errors concentrated around the resonance frequencies. 
In sharp contrast, the proposed approach is able to provide estimates with good accuracy regardless of the chosen local model order. This robustness is one of the main reasons why the proposed local modeling approach is superior to alternative parametric methods. 

\begin{figure}[ht]
    \centering
    \includegraphics{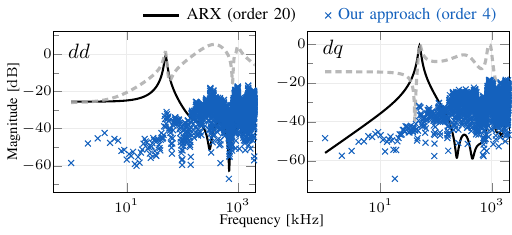}\vspace{-0.3cm}
     \caption{Error magnitude $|\hat{Z}_{dd}(j\omega) - Z_{dd}(j\omega)|, \; |\hat{Z}_{dq}(j\omega) - Z_{dq}(j\omega)|$ in case of noise corrupted measurements. The errors of $\hat{Z}_{qd}(j\omega)$ and $\hat{Z}_{qq}(j\omega)$ (not shown) exhibit the same behavior. The magnitude of the true responses (dashed gray) are overlaid to highlight the location of the resonances. }\vspace{-0.25cm}
    \label{fig:errors_with_noise}
\end{figure}

\begin{figure}
    \centering
    \includegraphics{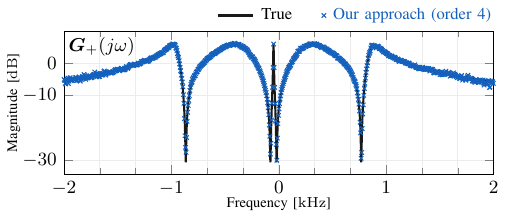}\vspace{-0.3cm}
         \caption{Magnitude frequency response of the complex TF $\bG_+$ 
         and an estimate  obtained using the proposed approach via a local model order 4, with noisy measurements (for clarity, only each 5th estimated frequency is shown).}\vspace{-0.25cm}
    \label{fig:errors_complex_TF}
\end{figure}

These observations become evident by inspecting the error plots in Figure \ref{fig:errors_with_noise}. The errors of the discrete-time parametric ARX method are concentrated at low frequencies around the resonances (true responses are overlaid in dashed gray). In contrast, the errors of the proposed approach become noticeable only at frequencies higher than 2 \si{\kHz}. Below this frequency, the estimates are accurate; see Figure \ref{fig:errors_complex_TF} where the estimates of $\bG_+$ are shown together with the true values. The observed worsening in the accuracy above 2 \si{\kHz} is due to the limitations imposed by the $LCL$ filter that reduces the SNR at higher frequencies. 

\renewcommand{\arraystretch}{1.1}
\begin{table}[b!]\scriptsize
    \centering
    \vspace{-3mm}
           \caption{Accuracy comparison over frequency band $[0,2]$ \si{\kHz},\\  \textit{with} measurement noise}
           \vspace{-0.2cm}
               \begin{tabular}{c c || c|c|c|c||c}
     \toprule
      \multicolumn{2}{c||}{\makecell{Method\\ (excitation time = 1 \si{\sec})}} & \multicolumn{4}{c||}{\makecell{Fit\%\\(rounded to a single digit)} } &   \multirow{2}{*}{\makecell{Relative\\ $H_\infty$ error}}\\
          \multicolumn{2}{c||}{RBS amplitude $\pm$ 0.05 p.u.} & $Z_{dd}$ &  $Z_{dq}$ & $Z_{qd}$ & $Z_{qq}$  &    \\ \hline
                \multicolumn{1}{c}{~}      &  \makecell{\rule{0pt}{2.5ex}order}  & & & &  \\[0.5em]
                \multicolumn{1}{c|}{\makecell{ \color{backgroundcolor}\textbf{Our approach} \\  \color{backgroundcolor}(non-parametric) }}  & \color{backgroundcolor}\bf  2  & \color{backgroundcolor}\bf 99.6 & \color{backgroundcolor}\bf  98.5 & \color{backgroundcolor}\bf 98.6 & \color{backgroundcolor}\bf 99.6  & \color{backgroundcolor}\bf 0.1229\\[-0.3em]
                \multicolumn{1}{c|}{~}                                                 &   \color{backgroundcolor}\bf4  &  \color{backgroundcolor}\bf99.7 &   \color{backgroundcolor}\bf98.8 &  \color{backgroundcolor}\bf 98.9 &   \color{backgroundcolor}\bf99.7  &  \color{backgroundcolor}\bf 0.1061\\
                \multicolumn{1}{c|}{~}                                                 &   \color{backgroundcolor}\bf6  &  \color{backgroundcolor}\bf99.7 &   \color{backgroundcolor}\bf99.0 &   \color{backgroundcolor}\bf99.0 &   \color{backgroundcolor}\bf99.7  &   \color{backgroundcolor}\bf0.0990 \\
                \multicolumn{1}{c|}{~}                                                 &   \color{backgroundcolor}\bf8  &  \color{backgroundcolor}\bf99.7 &   \color{backgroundcolor}\bf99.0 &   \color{backgroundcolor}\bf99.1 &   \color{backgroundcolor}\bf99.7  &   \color{backgroundcolor}\bf0.0957 \\
                \multicolumn{1}{c|}{~}                                                 &   \color{backgroundcolor}\bf10 & \color{backgroundcolor}\bf 99.7 &   \color{backgroundcolor}\bf99.0 &  \color{backgroundcolor}\bf 99.1 &   \color{backgroundcolor}\bf99.7  &   \color{backgroundcolor}\bf0.0936 \\
                &&&&&\\
               \multicolumn{1}{c|}{\multirow{3}{*}{\makecell{ discrete-time ARX\\ (parametric) }}}   &  2  &  48.6 & 10.3  & 11.5 & 52.0 &  0.8215\\
                \multicolumn{1}{c|}{~}                                                                     &  4  & 41.9 & 3.0  & 3.6  & 46.2 &  0.8183\\
                \multicolumn{1}{c|}{~}                                                                     &  6  & 42.3 & 3.8  &  6.5  & 49.5 &  0.8042\\ 
                \multicolumn{1}{c|}{~}                                                                     &  8  & 85.7 & 50.6 & 51.7 & 87.6 &  0.9286\\
                \multicolumn{1}{c|}{~}                                                                     &  10  & 97.6 & 87.2 & 85.8 & 97.9 &  0.9750\\
                \multicolumn{1}{c|}{~}                                                                     &  20   & 99.4  & 97.5 & 97.4 & 99.3 &   0.9186\\  
         \bottomrule
    \end{tabular}
     \label{tab:comparison_with_noise}
\end{table}
\renewcommand{\arraystretch}{1} \normalsize

\vspace{-0.5em}
\section{Conclusions}
We proposed a non-parametric method for identifying dynamic small-signal $dq$-asymmetric grid impedances using grid-connected converters. Our approach avoids assumptions about the grid's topology or structure and provides estimates over a wide frequency band using short, low-amplitude, non-periodic excitation. The key innovation lies in combining complex transfer function representations of asymmetric systems with local frequency-domain modeling techniques. This strikes a crucial balance between fully non-parametric methods, which require extended excitation times, and fully parametric methods, which are sensitive to the model order. Numerical simulations demonstrated the superior performance and robustness of our approach, with a measurement time of \mbox{1 \si{\second}} and a full frequency resolution of 1 \si{\Hz}. Our future work will cover the experimental validation of this method in more realistic settings,   identifying the converter admittance, error analysis, optimizing the excitation signal, and considering unbalanced and multi-converter scenarios.




%

\bibliographystyle{IEEEtran}
\bibliography{config/IEEEabrv.bib,config/literature.bib}

\newpage
\appendix

\!\!Define the continuous-time Fourier and convolution \mbox{integrals}
\begingroup\makeatletter\def\f@size{9}\check@mathfonts
\[
\F_a^b\{\bv\}(\omega) \!:=\!\!\int_a^b\! \bv(t) \eu^{-j\omega t} \diff t, \quad
(\bg\ast \bi)_a^b(t)  \!:=\!\! \int_a^b\! \bg(t-\tau) \bi(\tau) \diff \tau,
\]
\endgroup
and, for clarity, let us drop their arguments, $\omega$ and $t$, from the notation. We start by deriving the expressions for the $dq$-symmetric case. The general case is established similarly, considering the responses of $\bG_+$ and $\bG_-$ separately. 

\subsection*{The symmetric case}

First, note that $\bv(t) = (\bg\ast \bi)_0^t  + (\bg\ast \bi)_{-\infty}^0$
where $\bg$ is the impulse response of the stable and \textit{causal} TF $\bG$ and $t\in[0,T]$ with $T=NT_s$ where $N$ is the number of samples, and $T_s$ is the sampling time. Then
\[
\begin{aligned}
    \bV(\omega)   &:= \F_{-\infty}^\infty\{\shat \bv\} = \F_0^T\{\bv\}  \\  
                  &=  \F_0^T\{(\bg\ast \bi)_0^t\} +  \F_0^T\{(\bg\ast \bi)_{-\infty}^0\} \\
                  & =  \F_0^T\{(\bg\ast \bi)_0^\infty\} -\underbrace{\F_0^T\{(\bg\ast \bi)_t^\infty\}}_{= \, 0 \text{ by causality of $\bG$}} +  \F_0^T\{(\bg\ast \bi)_{-\infty}^0\} \\[0.5em]
                  &=   \F_0^\infty\{(\bg\ast \bi)_0^\infty\} -\F_T^\infty\{(\bg\ast \bi)_0^\infty\} +\F_0^T\{(\bg\ast \bi)_{-\infty}^0\}\\
                  &= \bG(j\omega) \bI(\omega) - \bV_{\!\text{fin}}(\omega) + \bV_{\!\text{init}}(\omega).
\end{aligned}
\]
Here, $\bV(\omega)$ and $\bI(\omega)$ are the spectra of the complex continuous-time signals $\shat \bv$ and $\shat\bi$ which are defined as being equal to $\bv$ and $\bi$ over the finite time interval $[0, T]$ and equal to zero elsewhere. In the third row, the second term is identically zero for a casual $\bG$ (because then $\bg(t) = 0 \; \forall t <0)$. The last equality follows from the Fourier transform convolution theorem, which implies $\F_0^\infty\{(\bg\ast \bi)_0^\infty\} = \bG(j\omega) \F_0^\infty\{\bi(t)\}$, and assuming $\bi(t) =0 \; \forall t>T$ (this assumption is benign because $\bG$ is causal); hence $\F_0^\infty\{(\bg\ast \bi)_0^\infty\} = \bG(j\omega) \bI(\omega)$. Lastly, we defined the terms  $\bV_{\!\text{fin}}(\omega):= \F_T^\infty\{(\bg\ast \bi)_0^\infty\}$, $\bV_{\!\text{init}}(\omega) :=\F_0^T\{(\bg\ast \bi)_{-\infty}^0\}$ as the spectral leakage terms.
\smallskip

The next step is to relate the continuous spectra $\bV(\omega)$  and $\bI(\omega)$ to the discrete ones obtained by the DFT in \eqref{eq:dft_spectrum}. The Dirac delta functions $\delta(t)$ constitute the classical tool. Recall that these generalized functions are defined using continuous test functions $\varphi$ via the identity $\int_{-\infty}^{\infty} \delta(\tau) \varphi(\tau) \diff \tau = \varphi(0)$, from which all properties of $\delta$ can be deduced. For example, the impulse train $\sum_{n=-\infty}^\infty \delta(t+nT)$, which plays an important role in sampling, is a periodic function with a period $T$, and can be expanded using a Fourier series as
\[
\sum_{n=-\infty}^\infty \delta(t+nT) = \frac{1}{T} \sum_{n=-\infty}^\infty \eu^{jn\omega_1t}, \qquad \omega_1 := \frac{2\pi}{T}.
\]
By convolving both sides of the last equation with $\shat \bv(t)$ we get
\[
\sum_{n = -\infty}^\infty \shat \bv(t+nT) = \frac{1}{T} \sum_{n=-\infty}^\infty \bV(n\omega_1) \eu^{jn\omega_1t}.
\]
In light of this, the Fourier transform symmetry theorem can be invoked to deduce the following relation between the samples $\{\bv(nT_s)\}$ and the continuous spectrum $\bV(\omega)$
\[
\sum_{n = -\infty}^\infty \bV(\omega+n\omega_s) = \frac{2\pi}{\omega_s} \sum_{n=0}^{N} {c(n)}{\bv(nT_s)} \eu^{-jnT_s\omega}.
\]
This is the celebrated Poisson's sum formula \cite[pg. 75, (3-87) \& Note 2]{papoulis1977signal}, where the finite limits of the sum on the right-hand side are because $\bV$ is defined by a finite-time Fourier integral over $[0,T]$, and the factor
\[
c(n) := \begin{cases}
    \frac{1}{2},\qquad n \in \{0,N\}\\
    1, \qquad \text{otherwise}.
\end{cases}
\]
is needed because of the discontinuity of $\shat \bv$  at the end points of the observation time window.

 Sampling the spectra uniformly over the unit circle, with a frequency resolution $\frac{2\pi}{T}$, leads to 
\[
\frac{2\pi}{\omega_s} \sum_{n=0}^{N} {c(n)}{\bv(t_n)} \eu^{-jnT_s\omega_k} = \sum_{n = -\infty}^\infty \bV(\omega_k+n\omega_s) 
\]
where $t_n := nT_s$, and $\omega_k :=\frac{2\pi k}{T}$, $k \in \{0, \dots, N-1\}$. Rearranging and using the definition in \eqref{eq:dft_spectrum}, we find that
\[
\bV\!_k = \frac{1}{\sqrt{N} T_s}\sum_{n = -\infty}^\infty \bV\!\left(\omega_n - n \omega_s\right) + \frac{\bv(0) - \bv(T)}{2\sqrt{N}}.
\] 
The same relation holds for $\bI_k, \bV_{\!\text{init},k}$, and $\bV_{\!\text{fin},k}$.  From this we get the model 
\[
\bV\!_k = \bG(j\omega_k) \bI_k + \bT(j\omega_k)
\]
where  $\bT(j\omega_k) = \bV_{\!\text{init},k} - \bV_{\!\text{fin},k} + \boldsymbol{\alpha}_k$ is the transient term with $\boldsymbol{\alpha}_k$ representing the aliasing effects.

\subsection*{The asymmetric case}
The starting point here is \eqref{eq:complex_model}, which may be re-written as
\[
\bv(t) = \bv_+(t) + \bv_-(t),
\]
with 
\[
\begin{aligned}
\bv_+(t) &= (\bg_+\ast \bi)_0^t  \;\,+ (\bg_+\ast \bi)_{-\infty}^0,\\
\bv_-(t) &= (\bg_-\ast \bi^\ast)_0^t  + (\bg_-\ast \bi^\ast)_{-\infty}^0,
\end{aligned}
\]
in which $\bg_+$ and $\bg_-$ are the impulse responses of causal $\bG_+$ and $\bG_-$, respectively. The above results, from the symmetric case, can be applied directly to $\bv_+(t)$ and $\bv_-(t)$ to show that the DFTs of $\{\bv_+(t_n)\}$ and $\{\bv_-(t_n)\}$ are
\[
\begin{aligned}
\bV\!_{+k} &= \bG_+(j\omega_k) \bI_k + \bT_+(j\omega_k),\\
\bV\!_{-k} &= \bG_-(j\omega_k) \bI_{(N-k)_N}^\ast + \bT_-(j\omega_k),\\
\end{aligned}
\]
The reversed and conjugated DFT spectrum of $\bi$ in the last equation arises because $\F_0^\infty\{(\bg\ast \bi^\ast)_0^\infty\} = \bG(j\omega) [\bI(-\omega)]^\ast$, after noticing that $\bI^\ast(\omega) := \F_0^\infty\{\bi^\ast(t)\} = [\bI(-\omega)]^\ast$. From this we directly get the model
\[
\bV\!_k = \bG(j\omega_k) \bI_k + \bG_-(j\omega_k) \bI_{(N-k)_N}^\ast + {\underbrace{\bT(j\omega_k)}_{=\bT_+(j\omega_k)+ \bT_-(j\omega_k)}}\vspace{-0.7em}
\]
 which is identical to \eqref{eq:dft_relation}.
 \medskip

Observe that the transient term $\bT(j\omega_k)$ carries two effects: 
\begin{enumerate}\renewcommand{\theenumi}{\roman{enumi}}
    \item  the spectral leakage captured by $\bV_{\!\text{init},k}$ and $\bV_{\!\text{fin},k}$; these two terms come from unforced decaying responses due to the initial and final conditions, respectively, and therefore they have the same poles as $\bG_+$ and $\bG_-$,
    \item  aliasing effect captured by $\alpha_k$; this term has infinite repetition of poles due to the folding of the spectrum.
\end{enumerate}

\end{document}